\documentstyle[twocolumn,epsfig,fleqn,amstex,rotating,multirow]{mn}
\newcommand{\hi}{\mbox{H\ {\footnotesize I}}}

\def\lsim{~\rlap{\raise 0.4ex\hbox{$<$}}{\lower 0.7ex\hbox{$\sim$}}~}
\def\gsim{~\rlap{\raise 0.4ex\hbox{$>$}}{\lower 0.7ex\hbox{$\sim$}}~}
\def\dd{{\rm d}}

\def\l0{L_\ast(0)}

\def\kb{k_{_{\rm B}}}
\def\s0{S_\ast(0)}

\def\ee{e^-e^+}
\def\omg0{\Omega_0}

\def\kms{{\rm km}\; {\rm s}^{-1}\;  {\rm Mpc}^{-1}}
\def\fh{f_{_{\rm heat}}}
\def\fion{f_{_{\rm ion}}}
\def\a2{\alpha^{(2)}}

\def\fcl{f_{\rm cl}}

\def\me{m_{\rm e} c^2}

\def\eph{E_{_{\rm ph}}}
\def\teph{{\tilde E}_{_{\rm ph}}}
\def\peph{{E'}_{_{\rm ph}}}

\def\EI{{E}_{_{\rm ion}}}

\def\sigmai{\sigma_{_{\rm ion}}}
\def\gE{\gamma_{_{_{\rm {E_{_{\rm ph}}} }      }}}

\def\tz{{\tilde z}}
\def\nen{N_{_{\rm enh}}}
\def\mpc3{\ {\rm Mpc^{-3}}}
\def\gpc3{\ {\rm Gpc^{-3}}}

\def\tcmb{T_{_{\rm CMB} }}

\begin{document}

\title[]
{ Neutralino  Annihilations and the Gas Temperature in the Dark Ages}
\author[Myers {\it \& Nusser}]{Zacharia Myers and  Adi Nusser\\\\
Physics Department, Technion, Haifa 32000, Israel
and the Asher Space Research Institute\\
}
\maketitle

\begin{abstract}
Assuming the dark matter is made entirely from neutralinos, 
we re-visit  the role of their  annihilation on the temperature of 
diffuse gas  in the high redshift universe. 
We consider  neutralinos  of particle mass $36 \;  {\rm GeV}$
and  $100\;  {\rm GeV}$, respectively. The former is able to produce 
$\sim 7$  $\ee$ particles per annihilation through  the fremionic 
channel, and the latter $\sim 53$ particles assuming a purely 
bosonic channel.    High energy $\ee$ particles
 up-scatter the Cosmic Microwave Background (CMB) photons into higher energies via the inverse-Compton scattering.
The process produces a power-law  $\ee$  energy spectrum  of index  
 $-1$  in the energy range of interest, independent of the initial energy distribution. The corresponding 
energy spectrum of the up-scattered photons 
is a power-law of index $-1/2$, if absorption by the gas is not 
included. The scattered photons photo-heat  the gas by releasing 
electrons which deposit  a fraction ($~14\%$) of their energy as heat 
into the ambient medium.  
For uniformly distributed neutralinos the heating  is insignificant.
The effect is greatly enhanced by the  clumping of neutralinos into dense haloes.
We use a time-dependent  clumping model which takes into 
account the damping of density fluctuations  on mass scales smaller than
$\sim 10^{-6}M_\odot$.
With this clumping model, the heating mechanism  boosts  the gas temperature above that of  the CMB after a redshift of $z\sim 30$. By $z\approx 10$ the 
gas temperature is  nearly 100 times its  temperature
when no  heating is invoked. Similar increase is obtained for
the two neutralino masses considered.   

\end{abstract}

\section {Introduction}
\label{sec:Introduction}

Early   formation of luminous objects (stars, galaxies, miniquasars)
in the universe   
is governed by an intricate  interplay between energetic (mechanical and radiative) feedback from these objects and the physical properties of the surrounding diffuse gas  (e.g. Benson et al. 2006).
Gas accretion onto dark matter (DM) haloes, and its cooling 
inside them,  depend on its  density and temperature  which  
are greatly affected by feedback from 
the forming luminous objects (e.g. Benson et al. 2006, Thomas \& Zaroubi 2007).
However, before the onset of the first generation of  luminous objects 
(e.g. Abel, Bryan, \& 
Norman {2002};  Bromm, Coppi, \&  Larson {2002}) 
and with the exception of coupling to the CMB, 
energetic sources affecting the state of the diffuse gas\footnote{We avoid  the term ``intergalactic medium (IGM)" to refer to the  gas 
at such high redshifts as it is unlikely that galaxies existed at those early times.}
could  only be associated with the dark matter through its decay, annihilation and 
direct collisions with ordinary matter.
Direct collisions  are relevant for super-heavy dark matter  (particle mass $\sim 10^5-10^{15} \; \rm GeV$ which does not violate bounds by ground experiments on the  interaction 
cross section with baryons (Albuquerque \& Baudis 2003).
However, the  cross section needed to heat the high redshift diffuse gas  is about 10 orders 
of magnitude larger than the upper bound derived from clusters of 
galaxies (Chuzhoy \& Nusser 2006).   
Various effects of dark matter decay and annihilation 
on  the gas at high redshift and  background radiations
have been studied (Sciama  1982;
Chen \&  Kamionkowski 2004;
Hansen \& Haiman 2004;
Kasuya, Kawasaki \& Sugiyama 2004;
Pieri \& 
Branchini 2004;
Pierpaoli {2004};
Padmanabhan \&  Finkbeiner 2005; 
Kasuya \&  Kawasaki 2006;
Mapelli et al. 2006;
Furlanetto et a. 2006;
Ullio et a. 2006; 
Zhang et al. {2006};
Freese, Gondolo \& Spolyar 2007; 
Ripamonti et al. 2007). 
Here we focus on  heating the diffuse gas as a 
result of  neutralino annihilations, taking into account their clumping  into small dense haloes (e.g. Green, Hofmann, \& Schwarz 2005) 
and a detailed analysis of energy deposition into the gas.
The main heating process is through 
inverse-Compton scattering (ICS) of  CMB photons by 
$\ee$ particles produced in the annihilations.    
The ICS process, involves the collision of a high-energy electron/positron  and a low-energy photon, with consequent production of a high-energy recoil photon and a corresponding decrease in electron energy, (Felten and Morrison, 1966). Once in contact with the CMB, these DM $\ee$ up-scatter the photons, shifting their energies to hard UV, X-ray and $\gamma$-ray levels. The heating effect from these up-scattered photons is only significant after DM clumping into (micro-) haloes of mass  $\sim 10^{-6} M_\odot$ has begun.
 The annihilation rate, which  is proportional to the
 DM density squared increases by orders of magnitude 
 inside these dense haloes. 
  

 The neutralino, $ \chi$,  is the lightest stable supersymmetric particle  (See Jungman, Kamionkowski, and Griest 1996 for a review). While much is still unknown about neutralinos because it is difficult to detect them directly,  they are considered to be Majorana fermions (and therefore are their own anti-particles,  $ \chi = \overline \chi$); thus, they self-annihilate and produce some combination of bosons, mesons, $e^- e^+$ pairs, and $\gamma$-rays (see Gunn et al. 1978 for the basic scenario). Baltz and Wai (2004) point out that a significant fraction of the power liberated in self-annihilation may go into high-energy $e^{+} e^{-}$ pairs. Monochromatic electrons (with   energy $\approx{M_{\chi}}$ ), coming from the direct channel  $\chi$ $\chi \rightarrow e^{+} e^{-}$, are generally suppressed (Turner and Wilczek 1990); electrons are then produced from the decay of the final heavy fermions and bosons. Low mass estimates for the neutralino such as 80 GeV or lower will not produce any bosons because they are too light. For estimates of $M_\chi={100 \; {\rm GeV}}$ or higher, DM annihilations produce bosons followed by a decay chain of subsequent particles including electrons.

The different composition of the  $\chi\overline\chi$ annihilation final states will in general affect the form of the final $\ee$  spectrum. Similar to Colafrancesco and Mele (2000), we consider two different cases: 1) a light mass neutralino, $M_{\chi}$=36 GeV, which yields mainly fermion pair production $\chi$ $\overline\chi \rightarrow f f $, (Turner and Wilczek 1990); 2) a heavier neutralino, $M_{\chi}$=100GeV annihilating into predominantly W (and Z) vector bosons, $\chi$ $\overline\chi \rightarrow W W (ZZ)$. A real situation will be mostly reproduced by either of the above two cases, or by a linear combination of the two (Colafrancesco \& Mele 2000).
     The $\ee$ energy spectrum arising from the $\chi$ $\overline\chi$ annihilation has been derived by various authors (Silk and Srednicki 1984, Rudaz and Stecker 1988, Ellis et al. 1989, Stecker and Tylka 1989, Turner and Wilczek 1990, Kamionkowski and Turner 1991, Baltz and Edsjo, 1998). Here, we adopt an approach similar to Rudaz and Stecker (1988) and  Kamionkowski and Turner (1991), that gave the analytical approximations of the electron source functions for models in which neutralinos annihilate mainly into fermions and vector bosons, respectively. Given the  source spectrum we can estimate the 'enhancement factor' $\nen$ defined as  the number of $\ee$ particles that are produced in a   single neutralino annihilation. 
     
Our  notation is as follows.
The scale factor  of the universe is $a(t)$, the Hubble function 
is $H(z)=\dot a/a$, the critical density is $\rho_{\rm crit}=
3H^2/(8\pi G)$. The mass densities (in units  of $\rho_{\rm crit}$) corresponding to dark matter  (i.e. neutralino), baryonic matter, dark  energy (cosmological constant), and curvature  are, respectively, $\Omega_{\rm M}$, $\Omega_{\rm B}$, 
are $\Omega_{\Lambda}$ and $\Omega_{\rm K}$. 
For the ratio, $h(z)=H(z)/H_0 $, where $H_0=H(z=0)$, we use 
\begin{equation}
h(z)=\sqrt{\Omega_{\rm M} (1+z)^3 +\Omega_{\rm K} (1+z)^2+\Omega_{\Lambda}}\; .
\end{equation}
We consider a flat universe, $\Omega_{\rm K}=0$, with 
$\Omega_{\rm M}=0.24$, $\Omega_{\rm B}=0.04$, $\Omega_{\Lambda}=0.72$
and $H_0=73\; \kms$ in accordance with Spergel et al. (2007).

The outline of the paper is as follows. In  \S2 we summarize the relevant  equations  used to obtain the enhanced temperature of the diffuse gas  at high redshifts. In \S3 we present the results  of the calculations. We conclude in  \S4.

\section {The equations}
The annihilation by-product particles, $\ee$, inverse-Compton scatter 
CMB photons to high energies. The up-scattered photons 
(herafter ICS photons) deposit part of their energy
into the diffuse gas through the process of photo-ionization. 
Our goal is to compute  the heating rate and the temperature of the  gas
as a function of time. 
The heating rate per baryon is given by 
\begin{equation}
{\cal H}(z)=4\pi \fh \int_{\EI} \frac{J(\eph,z)}{\eph} (\eph-\EI)\sigmai(\eph)\dd \eph \; ,
\label{eq:H}
\end{equation}
where $J(\eph)$ is the energy flux of the ICS photons in units of $\rm MeV \rm cm^{-2} \; sr^{-1}
\;s^{-1} \; MeV^{-1}$, $\EI=13.6\;  \rm eV$ is the \hi\ ionization energy, and 
$\sigmai=6\times 10^{-18}(\eph/\EI)^3\rm cm^{2}$ for $\eph>\EI$ and zero otherwise.  The factor $\fh$ is the fraction of 
energy of released electrons which is deposited as heat by 
collisions with the gas particles (atoms and free thermal electrons). This factor 
depends strongly on the ionization fraction of the gas, ranging from 
almost unity for nearly fully ionized gases to 
$\fh\approx 0.135$ for an ionization fraction of $10^{-4}$ (Shull \& Van Steenberg 1985). 
We ignore the  mild dependence of $\fh$ on the electron energy
and  consider its lower limit values  which are obtained 
for very high electron energies.   
The residual ionized fraction from the epoch of recombination is $
\approx 2 \times 10^{-4}$ (e.g. Peebles 1993), which gives the value $\fh=0.144$ 
(Shull \& Van Steenberg 1985) adopted in this work. 
The ICS energy flux, $J(\eph)$,  is  given by  (e.g. Haardt \& Madau, 1996),
\begin{equation}
J(\eph,z)=\frac{1}{4\pi} \frac{c}{H_0}\int_{z}^{z_{\rm rec}}\frac{\dd \tz}{(1+z)h(z)}
\epsilon({\teph},\tz) e^{-\tau(\eph,z,\tz ) }
\label{eq:JJ}
\end{equation}
where $c$ is the speed of light, ${\teph}=\eph(1+\tz)/(1+z)$,
 and $\epsilon$ is the emissivity (in units of MeV per $\rm s\; MeV \; cm^3$) of ICS radiation per 
unit volume comoving with respect to  an observer at redshift $z$, i.e., the emissivity 
per unit proper volume is  $(1+z)^3/(1+\tz)^3 \epsilon$.
In the above expression,  no contribution from 
$\ee$ generated from annihilations before the epoch of recombination
at  $z_{\rm rec}\approx 1000$ is taken into account. The reason is that
the mean free path of such $\ee$ particles is so small that they 
are thermalised with the cosmic plasma before they escape to lower redshifts.  
The optical depth for absorption is
\begin{equation}
\tau(\eph,z,\tz )=\frac{c}{H_0}\int_z^{\tz}\frac{\dd z'}{(1+z')h(z')}\sigmai(\peph)
 n_{_{\rm H}} (z')  
\end{equation}
where $n_{_{\rm H} }(z') \approx 1.8 \times 10^{-7} {\rm cm}^{-3}(1+z')^3$
 is the proper \hi\ density at redshift $z'$ and
$\peph=\eph(1+z')/(1+z)$. Here we consider absorption only by 
\hi\ .

The ICS emissivity depends on the CMB temperature $\tcmb(z)\approx 2.73 (1+z) {\rm K}=
2.3\times 10^{-10} (1+z) \;{\rm MeV}/\kb$ (where $\kb$ is Boltzmann constant),
its energy density $U=2.6 \times 10^{-7} (1+z)^4 \; \rm MeV\;  cm^{-3}$, and the number density  
of $\ee$ as a function of energy. 
The expression for $\epsilon(\eph,z)$ as a function of the photon 
energy is (Felten \& Morrison 1966),
\begin{equation}
\epsilon(\eph,z) =  \frac{P (\gE,U)  N(\gE) }{ 7.2(\kb\tcmb) \gE}\; .
\label{eq:em}
\end{equation}
In this expression, it is assumed that only electrons/positrons with a Lorentz 
factor of
\begin{equation}
 \gE =\Big(\frac{\eph}{3.6k\tcmb}\Big)^{1/2}
 \label{eq:ge}
\end{equation}
 contribute to the ICS flux of photons at  energy $\eph$. Further, $P(\gamma,U)$ is the   inverse-Compton power scattered by one electron and is given by (Felten \& Morrison 1966)
\begin{equation}
\label{eq:P}
P (\gamma,U) = \frac{4}{3} \sigma_{\rm T} c \gamma^2 U\; ,
\end{equation}
where $\sigma_{\rm T}=6.65\times 10^{-25} \; {\rm cm}^2$ is the 
Thomson cross-section, and $N(\gamma)\dd \gamma $ gives the comoving (with respect to an observer 
 at redshift $z$) 
 number density of $\ee$ particles with Lorentz factors in the range 
 $\gamma \rightarrow \gamma+\dd \gamma$.

\label{sec:Dark Matter Candidates}
\subsection {The electron energy distribution}
\label{sec:eed}

The energy distribution function $N(\gamma,z)$ of $\ee$ particles is determined by the energy losses they incur 
through ICS with the CMB.
For simplicity, we assume here that all electrons are produced with 
the same energy $E_0$, corresponding to a Lorentz factor of $\gamma_0=E_0/m_{\rm e} c^2$ ($\me$ is the electron mass). Although the generalization to a general initial source spectrum 
is straightforward, we will see that the evolved 
spectrum is insensitive to the details of the initial spectrum.

The production rate of electrons per unit volume (which is comoving with  an observer at redshift zero\footnote{This rate can be transformed 
to a unit volume which is comoving with observers at redshift $z_1$ by 
a multiplication with $(1+z_1)^3$.}) is
\begin{equation}
 \frac{\dd N_{{e^-e^+}}}{\dd t}=\nen\;  (n_{\chi}(z))^2  \langle\sigma v\rangle (1+z)^{-3}\; ,
 \label{eq:eprod}
 \end{equation}
where $n_\chi(z)=\Omega_{\rm M} \rho_{\rm crit}/M_{\chi}\propto (1+z)^3$
is the proper neutralino number density at redshift $z$
 and $\nen$ is 7.6 and 53.2 for electrons produced from fermionic and bosonic decay chains, respectively.  
The annihilation cross section is 
$\langle\sigma v\rangle= 2\times 10^{-26} {\rm cm}^3 {\rm s}^{-1}$ ( e.g. Finkbeiner 2005).

 The energy loss rate of an electron as a result of scattering with the CMB
is given by 
\begin{equation}
\frac{\dd \gamma}{\dd t}=-\frac{P(\gamma) }{m_{\rm e} c^2}\; .
\end{equation}
Taking  $P$ from equation ({\ref{eq:P}) yields, 
\begin{equation}
\frac{\dd \gamma}{\dd t}=-A \gamma^2 (1+z)^4\; ,
\label{eq:eloss}
\end{equation}
where $A=2.7 \times 10^{-20} {\rm s}^{-1}$.
The $\ee$ particles can also lose energy by direct collisions with the gas. 
The corresponding  energy loss rate is 
$\dd \gamma/\dd t|_{_{\rm coll}}=\gamma c n_{_{\rm H}} \sigma_{_{\rm eH}}=
3\times \ln(\gamma) 10^{-15}[(1+z)/10]^3 {\rm s}^{-1}$ where
$\sigma_{_{\rm eH}}$ is the cross section for direct collisions
(Shull \& Van Steenger 1985).  By comparing with 
(\ref{eq:eloss}) we see that  direct collisions are more important than ICS losses for 
Lorentz factors $\gamma<\gamma_{_{\rm coll}}\approx 4(\ln \gamma)^{1/2}[(1+z)/ 10]^{-1/2}$.
The fraction of energy in $\ee$ particles with $\gamma <\gamma_{_{\rm coll}}$ relative to the total energy produced by annihilations is 
$\sim \gamma \me /M_{\chi}< 10^{-4}$. 
This is to be compared with the fraction of energy 
in electrons released by ICS photons. This fraction is $\sim 5\%$ as
we shall see below.  
Thus heating through direct 
collisions of $\ee$ particles with the gas is negligible.

The comoving number density (with respect to an observer at redshift zero) 
$N(\gamma, z) $ is related to the production rate in equation (\ref{eq:eprod})
by 
\begin{equation}
N(\gamma,z)\dd \gamma=\frac{\dd N_{\ee}(z_{\rm p})}{\dd t} \dd t_{\rm p}\; ,
\label{eq:cont}
\end{equation}
where  $z_{\rm p}(\gamma)$ is the redshift at which an $\ee$ particle  with 
Lorentz factor $\gamma$ (currently present at $z$) is produced and  $\dd t_{\rm p}$ is the difference in the production times 
of electrons with $\gamma$ and $\gamma+\dd \gamma$. The redshift 
$z_{\rm p}$ 
 is  determined  by  equation (\ref{eq:eloss}) with the initial condition that 
the electron is  produced with $\gamma_0$.
Together with the   equations (\ref{eq:eprod},  (\ref{eq:eloss}),
the relation  (\ref{eq:cont}) 
 yields,
\begin{equation}
N(\gamma,z)=  \frac{N_0}{(1+z_{\rm p}) \gamma^{2}}\; 
 \label{eq:nz}
\end{equation}
where
\begin{equation}
N_0=\frac{\nen}{A} n_\chi^2 \; 
 \langle\sigma v\rangle\; .
\label{eq:n0}
\end{equation}
At high redshifts, $(1+z) \propto t^{-2/3}$ and the energy loss 
equation (\ref{eq:eloss}) gives  $z_{\rm p} (\gamma)$ in terms of 
\begin{equation}
(1+z_{\rm p})^{5/2}=\frac{5}{2}\Omega_m H_0 \Big(\frac{1}{A\gamma} - \frac{1}{A\gamma_0} \Big) + (1+z)^{5/2}\; .
\end{equation}
For the relevant $\gamma $  values,  this relation 
gives $z_{\rm p}$. 
For an electron production energy of about $E_0 \lsim \; 1 \; {\rm GeV}$ we have $H_0/(A\gamma_0)\sim 0.04$
which is much smaller than $(1+z)^{5/2}$ at high $z$. Therefore, 
as long as 
 $\gamma$ is such that   $H_0/(A\gamma)\ll (1+z)^{5/2} $ is satisfied  we get  $z_{\rm p}$ very close to $z$ and 
$N(\gamma) \sim \gamma^{-2}$. Note that apart from setting an upper limit on the 
 $\gamma$, the value of $\gamma_0$  affects neither the shape nor the amplitude 
 of $N(\gamma) $ below that limit.
For $ H_0/(A\gamma)\gg (1+z)^{5/2} $ we have $z_{\rm p} \propto \gamma^{-2/5}$ which 
gives $N(\gamma) \propto \gamma^{-8/5}$, instead of $\gamma^{-2}$.
However, at $z\gsim 10$, for  $\ee$ energies spanning 
the range $1-10^{3}\; \rm  MeV$, we find that $z_{\rm p}=z$ to within a
few percent.

So far, we have assumed that  neutralinos 
are distributed  uniformly with a number density of $n_\chi=\bar \rho/M_{\chi}$, where $\bar \rho=\Omega_{\rm M}\rho_{\rm crit}$ is the mean background density.
The annihilation rate at a point $\bf x$ in space is proportional to the square of the denisty, $\rho({\bf x})$. 
Hence,  clumping of neutralinos increases the mean annihilation rate  by 
the  factor  $\fcl=\langle\rho({\bf x})/\bar \rho\rangle_{_{\bf x}}^2$
which is the variance of mass density  fluctuations. 
Fluctuations in the neutralino distribution are damped below a mass 
scale of $M_{\rm d}\sim 10^{-6}M_\odot$ because of neutralino collisions and 
and free streaming (Green, Hofmann, \& Schwarz 2005). 
We  define a nonlinear clustering scale, $M_{\rm nl}$ by 
$\sigma_\delta(M_{\rm nl},z)=1$, where $ \sigma_\delta(M,z)$ is the rms value of density fluctuations 
smoothed with a top-hat window on the mass scale $M$.
For sufficiently high redshifts we 
have $\sigma_\delta(M_{\rm nl},z)\ll M_{\rm d}$ and 
no significant clumping is obtained.  
The mass fluctuations over a given mass scale grow with time and 
hence, $M_{\rm nl}$ will exceed $M_{\rm d}$ at some redshift. 
When this occurs, neutralinos begin to cluster into 
the first generation of haloes  (of mass $>M_{\rm d}$) and the clumping factor, $\fcl$,  becomes
significant.   
To compute  $\fcl$ at any redshift, we proceed as follows.  
 If the fraction of mass in haloes is $F_{\rm h}(z)$ 
then $\fcl=(1-F_{\rm h}) + F_{\rm h} S$ where
$S$ depends on the density profile of the halo.
Using the parametric form which matches the density profiles in 
simulations of first generation  haloes (Diemand, Moore \& Stadel 2005)
gives $S=1200$.   Further, we 
estimate  $F_{\rm h}(z)$ according to  the  Press-Schechter 
(Press \& Schechter 1974) cumulative mass function 
 for the fraction of mass in haloes with mass larger than $M_{\rm d}$ 
 where we assume that $M_{\rm d}=M_{\rm nl}$ at $z=60$ (e.g. Pieri et al. 2007; Green, Hofmann \& Schwarz 2005). 
 The clumping factor is plotted in figure (\ref{fig:fcl}).

\begin{figure}
\centering
\mbox{\epsfig{figure=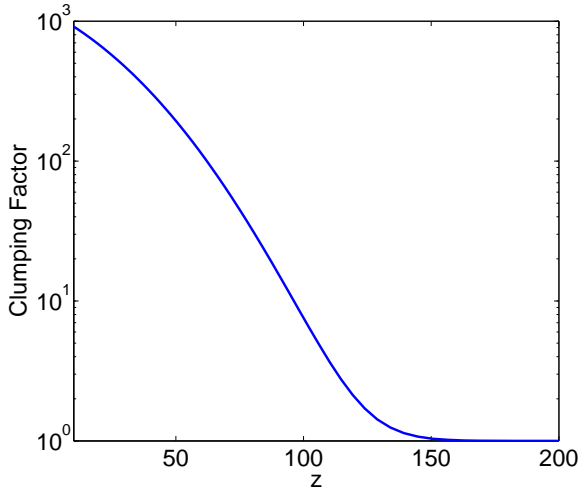,height=2.7in}}
\vspace{0.3cm}
\caption{The neutralino clumping factor, $\fcl$,  as a function of redshift.}
\label{fig:fcl}
\end{figure}

\section {Results}

\subsection{The electron spectrum}
In figure (\ref{fig:NE}) we plot, $N/m_{\rm e} c^2$, i.e.  the number density of  
$\ee$ particles  per unit energy as a function of the particle energy.
The upper and lower panels correspond to neutralino masses 
of  $36\; \rm  GeV$ and $100\; \rm GeV$, respectively. 
In each panel the spectra at redshifts, $z= 50$, 30, 20 and 10
are shown. Only annihilations at redshifts $z<1000$ are considered.  
A  slope of $-2$ describes all curves extremely well. The transition 
to $-8/5$ is almost undetectable down to energies close to the 
rest mass of the electron. For $M_\chi=36\rm \; GeV $
the fermionic channel is applicable and for  $M_\chi=100\rm  \; GeV $
we assume a purely bosonic channel.
While electron number density scales as $1/{M_\chi^2}$ (see eq.~\ref{eq:n0}), the light ($M_\chi={36\; \rm  GeV}$) and heavy ($M_\chi={100\;  \rm GeV}$) neutralinos
produce similar  heating rates. 
 This is because the heavy neutralino 
 is assumed to produce $\ee$ through the bosonic decay chains which  yield a larger enhancement factor ($\nen\sim 53$) than the 
 fermionic channel ($\nen\sim 7$) responsible  for $\ee$ 
 production through annihilations of the lighter particle.

\subsection{The spectrum of ICS photons}

Given the energy distribution $N(\gamma)$,  the spectrum of the ICS photons can readily be computed at any redshift using the expression (\ref{eq:JJ}). In figure   (\ref{fig:ICP}) we show, $J(\eph)/\eph$, i.e. number flux of ICS photons. The energy range extends to 
 much higher energies, but we plot the flux only for the range which is  relevant for heating the gas. 
 Substituting the relations (\ref{eq:nz})
and  (\ref{eq:em}) in ({\ref{eq:JJ}) we obtain $J/\eph\propto \eph^{-1.5}$ 
for $\tau=0$. This behaviour is clearly seen in the curves corresponding to $\tau=0$. 
When photo-ionization losses are included,  a deep dip occurs at 
  at $1.36\times10^{-5}$ MeV -   the ionization energy threshold for hydrogen.  The dip does not get down to zero due to ICS photons
 up-scattered during  a time period of $\delta t_{\rm ion}= 1/(c n_{_{\rm H}} \sigma_0) \sim 
 (1+z)^{-3} 1.5 \;\rm Myr$ preceding 
 $z$, the redshift at  which $J$ is given. 
 Photo-ionization losses disappear at sufficiently 
 high photon energies because of the decrease of 
  the ionization cross-section with increasing energy ($\sigmai\propto \eph^{-3}$).  
 The amplitude of the spectrum is larger at higher redshifts.
  This is to be expected given the hotter, denser, CMB at higher redshifts. 
The dip extends from $\EI $  up to $1 \; \rm keV$.
This is also the range of the kinetic energies of the released primary electrons. A  
detailed calculation gives a primary electron mean energy of $\sim 50 \;  \rm eV$.

\begin{figure}
\centering
\mbox{\epsfig{figure=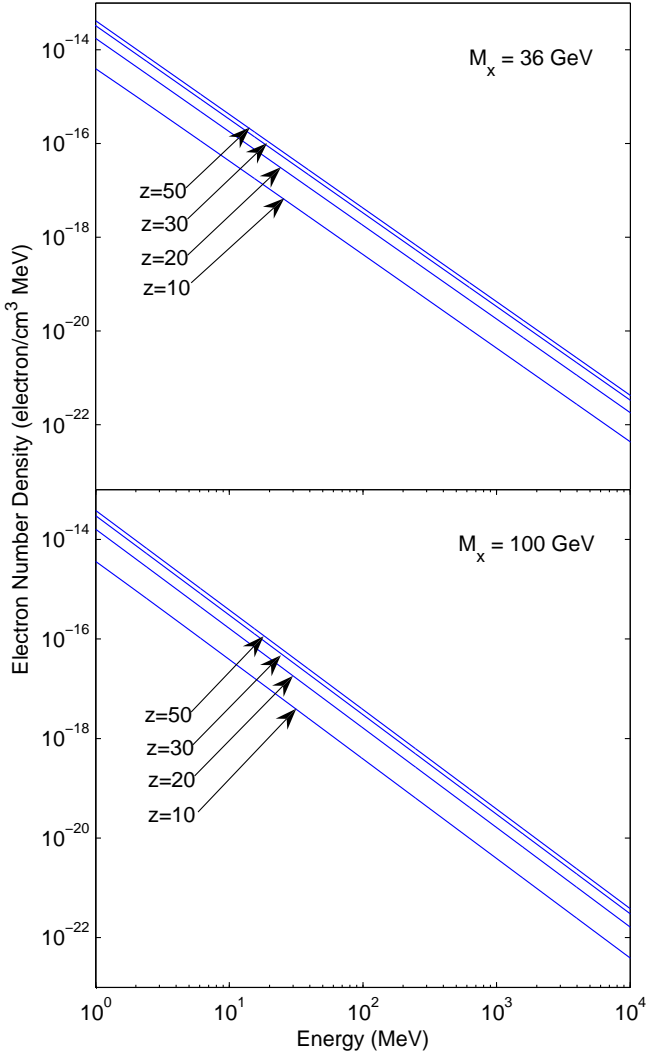,height=5.7in}}
\vspace{-0.6cm}
\caption{Spectra of  $\ee$ particles, resulting from  
IC losses to CMB photons, for neutralino masses of 
$M_\chi=36\;  \rm GeV$  (top) and $100\; \rm GeV$ (bottom). Each panel shows spectra computed for 4 redshifts as indicated. Only annihilations 
below redshift 1000 are considered. }
\label{fig:NE}
\end{figure}

\begin{figure}
\centering
\mbox{\epsfig{figure=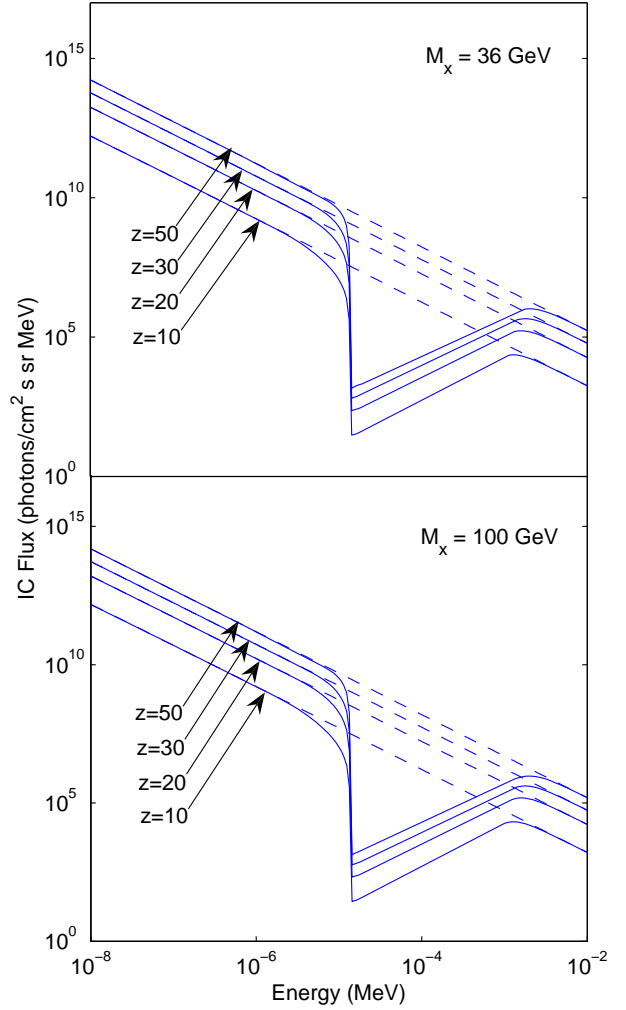,height=5.7in}}
\vspace{-0.6cm}
\caption{ Spectra of ICS photons at difference redshifts,  for  $M_\chi={36 \; \rm GeV}$ (top)  and $M_\chi={100\;  \rm GeV}$ (bottom). The solid line represents the ICS spectrum accounting for photon loss to \hi\ ionizations; hence the noticeable dip. The dashed line is the spectrum without accounting for the loss to ionization.}
\label{fig:ICP}
\end{figure}

\subsection{The heating rates and the gas temperature}
The gas temperature is governed by the energy equation,
\begin{equation}
\frac{\dd T }{\dd t}={\cal H}-2\frac{\dot a}{a}T+ \frac{T}{t_{_{\rm Compt}}}\; ,
\label{eq:eeq}
\end{equation}
where $\cal H$ is  the heating rate by ICS photons, 
as given in (\ref{eq:H}), the  second term on the 
right is adiabatic cooling 
 of perfect gas at mean cosmic density,   and
 the last term describes 
energy transfer between the CMB and the gas via Compton     
scattering of CMB photons with  thermal free electrons.  The time-scale, $t_{_{\rm Compt}}  $ (Peebles 1968) is,
\begin{equation}
t_{_{\rm Compt}}=\frac{1161.3(1+y^{-1})}{(1+z)^4[1-\tcmb(z)/T]} {\rm Gyr}\; .
\end{equation} 
In this expression,  
the ionization fraction, $y$,   obeys 
the equation
\begin{equation}
\frac{\dd y}{\dd t}={\cal I}-\alpha(T)y^2 n_{_{\rm H}}(z)\; ,
\label{eq:yeq}
\end{equation}
where $\alpha= 6.3\times 10^{-11}T^{-1/2}(T/10^3)^{-0.2}/(1+(T/10^6)^{0.7})\; {\rm s}^{-1}$  ($T$ in K) is the recombination rate and 
\begin{equation} 
{\cal I}=4\pi  \int_{\EI} \frac{J}{E}\left[1+
\fion \left(\frac{E}{\EI}-1\right)\right]
\sigmai(E)\dd E \; ,
\end{equation}
which includes contributions from two processes: direct ionization by ICS photons
and from secondary  electrons created by the  
primary energetic electron released by the first process. 
Ionizations by secondary electrons  consume a fraction  $\fion$ of the energy of primary  electrons. There is a dependence on $\fion$ on 
$y$, but is rather weak at the relevant $y$ values we 
get here.  We simply take  $\fion=  0.36$ corresponding to 
an ionization fraction of $2\times 10^{-4}$ (Shull \& Van Steenberg 1985).

The top  panel of figure (\ref{fig:rates}) shows  the  heating  rate of the  gas as a function of redshift. 
The rates include  clumping according to the scheme described at the end of \S\ref{sec:eed}.  
It is interesting to compare this rate to the total energy rate  produced  by  dark matter annihilations, i.e. 
$\langle\sigma v\rangle[n_\chi(z)]^2 M_\chi \fcl(z)/n_{\rm H}(z)$.  This ratio gives the heating  efficiency of the ICS photons.  
The ratio, plotted in figure (\ref{fig:ratiorates}), is fairly constant 
for $z>30$ at the level of $4\%$, but increases up-to $7\% $ towards $z=10$. 
The ionization rates, seen in the bottom panel, 
are too small to significantly increase the ionization fraction 
above its residual value from the epoch of recombination. 
A numerical solution of the ionization equation (\ref{eq:yeq})
 yields a maximum ionization fraction 
of $\sim 10^{-3}$ which is  obtained at the lowest redshifts 
 considered here, $z\sim 10$.

Given the heating and ionization rates, we numerically solve equations 
(\ref{eq:eeq}) and (\ref{eq:yeq}) to obtain the gas temperature 
as a function of redshift. 
Figure (\ref{fig:temp}) shows numerical solutions with the  initial 
condition $T=\tcmb$ at $z=500$.
The dash-dotted  curve representing $T(z)$ with no heating 
is obtained assuming $y=2\times 10^{-4}$ at all redshifts. 
Nonetheless, Compting coupling (heating in this case) plays no role 
at the plotted redshift range as the curve with no heating declines 
as $(1+z)^2$ as expected from adiabatic cooling alone. 
The CMB temperature (dashed curves) follows $(1+z)$ and it is above the 
gas temperature when  heating is not invoked. 
The solid curves in the two top panels  are the temperature 
when ICS heating is included. Both neutralino masses 
yield similar gas temperatures.
The rise in the gas temperature at $z>30$ is significant, but is not enough 
to bring the gas above $\tcmb$.  
At $z<30$ the heated gas  temperature
exceeds the  $\tcmb$.

\begin{figure}
\centering
\mbox{\epsfig{figure=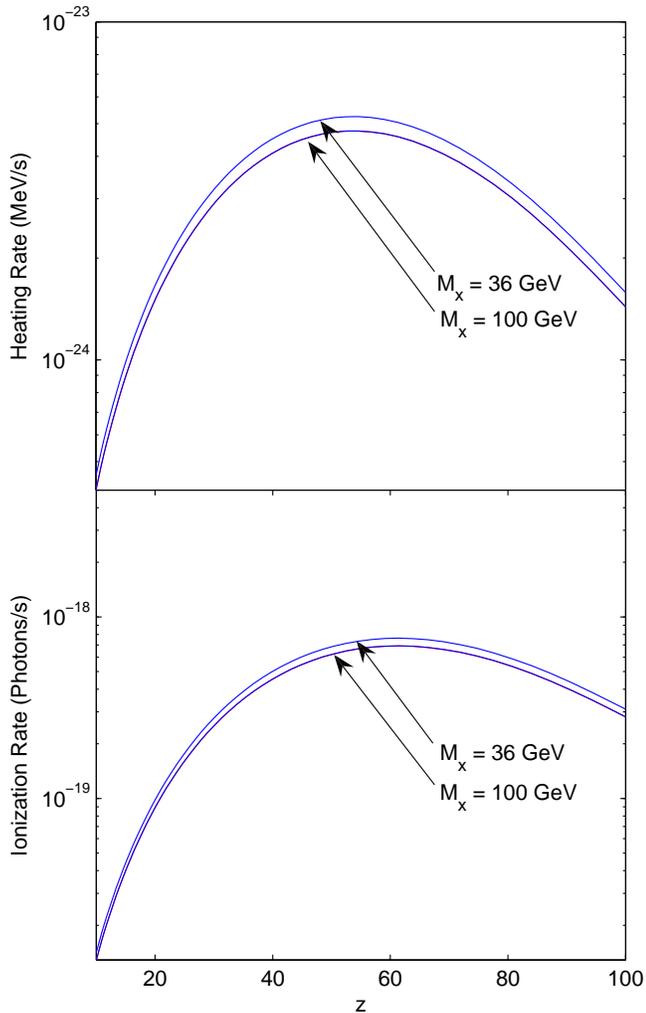,height=5.7in}}
\vspace{-0.6cm}
\caption{The heating (top panel) and ionization (bottom) rates  as a function of redshift for   $M_{\chi}$ = 36 GeV and $M_{\chi}$ = 100 GeV
as indicated in the figure. }
\label{fig:rates}
\end{figure}

\begin{figure}
\centering
\mbox{\epsfig{figure=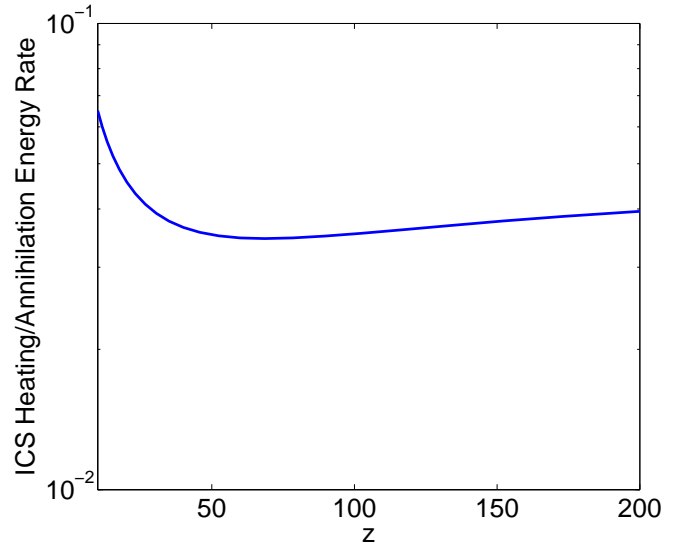,height=3.in}}
\vspace{0.3cm}
\caption{The ratio of the gas heating  by ICS photons to 
the  rate of total energy released in DM annihilations.}
\label{fig:ratiorates}
\end{figure}

%
\begin{figure}
\centering
\mbox{\epsfig{figure=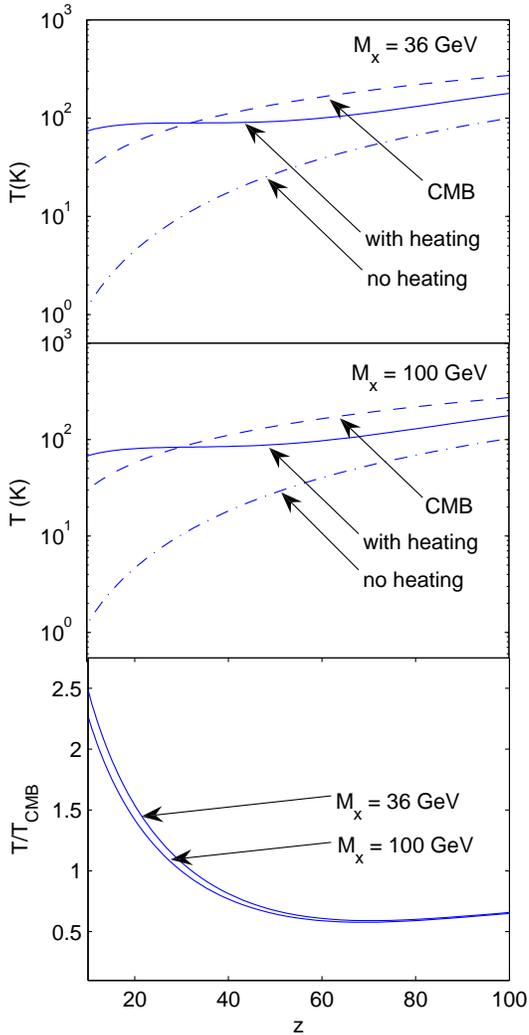,height=5.7in}}
\vspace{-0.6cm}
\caption{Top and middle panels show temperature of diffuse gas  
with and without heating, as a function of redshift. Plotted also is  the temperature of the CMB. The  bottom panel shows the ratio of the heated gas temperature relative to that of the CMB.}
\label{fig:temp}
\end{figure}

\section {Conclusions}

The heating of the gas by neutralino annihilations is   
 mediated by CMB photons up-scattered by collisions with 
 energetic  $\ee$ particles generated as a by-product of the annihilation.
Significant heating is obtained due to the clumping  neutralinos into  haloes.
Since the heating rate is directly proportional
to the clumping factor, comparison between the figures 
(\ref{fig:fcl}), (\ref{fig:rates}) and (\ref{fig:temp}) 
reveals that only negligible heating could be achieved if clumping 
is not included. 
The simple clumping model we adopt here could be improved to account for the dependence of halo profile on redshift and 
mass. These effects may enhance the clumping factor, yielding 
a more significant heating rate. 
A more precise model of the clumping factor could 
be achieved by additional high resolution simulations of 
the early stages of neutralino clustering.

It is interesting to see how patchy the ICS heating is 
when the mass fraction in haloes is small. 
Electrons/positrons produced in a single halo, will travel some distance  away from their origin before their energy 
becomes low enough so that the corresponding  ICS photons 
are capable of ionizing \hi . 
The degree of patchiness could be assessed by a  comparison of this distance with 
  the mean separation between haloes. 
According to (\ref{eq:ge}), an electron (or positron) with Lorentz factor $\gamma$
up-scatters a CMB photon to energy (in eV)  $\eph =
8.2\times 10^{-4}\gamma^2(1+z)$.
Therefore, $\gamma \sim  40 $ is needed to bring a $z=10$ CMB photon at  to $\sim \EI=13.6\; {\rm eV}$, the  \hi\ ionization threshold.   
The time it takes an electron to 
lose energy from its initial value $\gamma_0\gg 40$ down to $\gamma \ll \gamma_0 $   is
$[\gamma A (1+z)^4]^{-1}$ which gives  2.3 Myr for our photon of 
$\gamma \sim 40$ at $z\sim 10$. The comoving distance travelled by 
the electron during this time is $\sim 10\;  \rm Mpc$. This is 
 huge compared to the mean separation,  $3 F_{\rm h}^{-1/3} \;  \rm pc$,
 between haloes of mass $\sim 10^{-6}M_\odot$, unless the mass fraction,
 $F_{\rm h}$ 
  is extremely small.  
Therefore,  the ICS photons form a uniform 
background.

The temperature increase could  be significant for determining 
the onset of galaxy formation. 
 A  temperature
$T$ corresponds to a potential depth of a halo of mass, 
$M=\frac{20}{1+z}(\frac{T_{\rm v}}{10^5})^{3/2} 10^9M_\odot$ for
  $\Omega_{\rm M}\approx 1$ as is the case at high redshifts. 
A gas  at temperature $T $ 
could only be gravitationally trapped by haloes with  $T_{\rm v}$ 
 greater than  $ T$. 
For a gas cooling adiabatically without any heating mechanism 
we get $T=10{\rm K}$ at $z\approx 20$. Therefore,  gas can collapse  onto haloes of mass $> 10^3 M_\odot$.
At this redshift,
the heating mechanism described here, boosts the temperature of the gas by 
a factor of 10 at $z\approx 20$ which raises  its mass threshold for 
 collapse to  
$3\times 10^4 M_\odot$. This change in the mass threshold could 
result in a significant delay in the onset of star/galaxy formation. 

The temperature increase could be relevant for observations of 21 cm 
radiation from \hi\ at high redshifts. 
The 21 cm differential   brightness temperature
is 
$\delta T_{\rm b}\approx 16 {\rm mK} (1+\delta) [(1+z)/10]^{1/2}(T_{\rm s}-\tcmb)/T_{\rm s}$
where $\delta $ is the gas density contrast.
Collisional  coupling (e.g. Field 1959) of the kinetic and spin
temperatures  of the gas  could boost the latter above  $\tcmb$.
An  estimate of this coupling  gives $(T_{\rm s}-\tcmb)/T_{\rm s}\approx 0.1$ at $z\approx 20$ for a density contrast of unity
which gives  $\delta T_{\rm b}\approx  4 \rm mK$. This  is not far  
from  the sensitivity of planned 21 cm experiments.

\section*{Acknowledgements}
This research is supported by the German-Israeli Science Foundation for Development 
and Research and by the Asher Space Research fund.
ZM wishes to thank the Israeli Ministry of the Absorption of Science for providing the research funds necessary to support this work.

\end{document}